\documentclass{article}
\usepackage{hyperref}

\hypersetup{%
  colorlinks=true,
  linkcolor=blue,
  urlcolor=blue,
  citecolor=black
}
\usepackage[margin=1.5in]{geometry}

\usepackage[T1]{fontenc}
\usepackage[utf8]{inputenc}

\usepackage{graphicx}
\usepackage{fancyhdr}

\usepackage{utopia}
\usepackage{cite}

\fancypagestyle{broad_logo}{\fancyhf{}\fancyfoot[L]{\includegraphics[width=.4\linewidth]{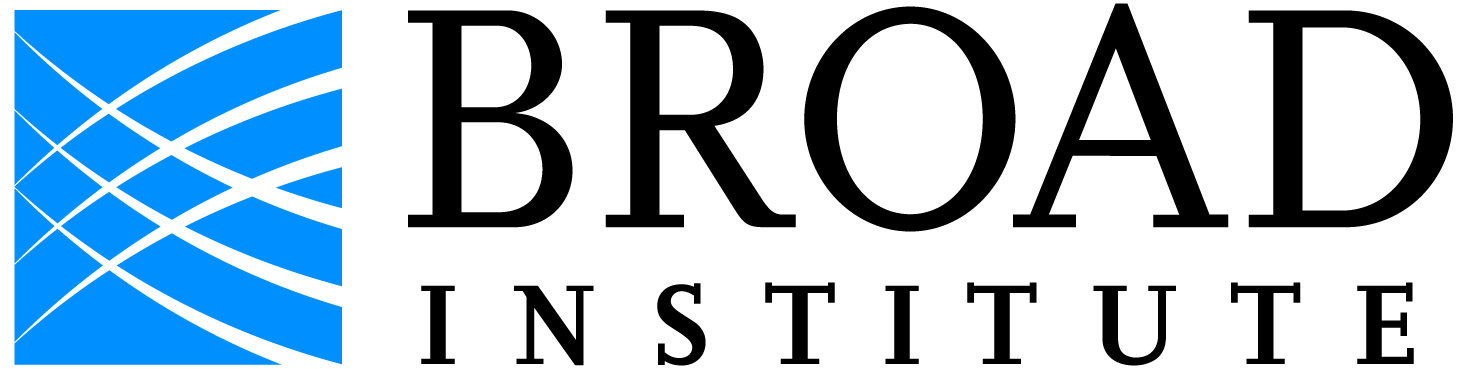}}}

\fancypagestyle{contact_info}{\fancyhf{}\fancyfoot[L]{Corresponding author: gsarma@broadinstitute.org}}

\title{Formal Methods for the Informal Engineer: Workshop Recommendations}
\date{}
\begin{document}

\newpage
\thispagestyle{empty}
\mbox{}
\newpage
\maketitle
\thispagestyle{broad_logo}

\pagebreak

\begin{center}
{\Large Workshop Leadership and Contributors}
\end{center}
\thispagestyle{contact_info}
\section*{Program Committee}
\begin{tabular}{l l}
Gopal Sarma & Broad Institute of MIT and Harvard \\
James Koppel & MIT CSAIL \\
Gregory Malecha & BedRock Systems \\
Patrick Schultz & Broad Institute of MIT and Harvard \\
Eric Drexler & James Martin School of Oxford University \\
Ramana Kumar & DeepMind
\end{tabular}

\section*{Tutorial Leaders}
\begin{tabular}{l l}
Cody Roux \hspace{.92cm} & Draper Labs \\
Philip Zucker & Draper Labs
\end{tabular}

\section*{Speakers}
\begin{tabular}{l l}
Kathleen Fisher & Tufts University \\
Gregory Malecha & BedRock Systems \\
Hillel Wayne & Windy Coast Consulting \\
Matthew Mirman & ETH Zurich 
\end{tabular}

\pagebreak


\section{Introduction}
In 2021, a workshop was convened at the Broad Institute of MIT and Harvard to explore potential applications of formal methods and programming language theory to software platforms being developed in the life sciences. The vision to host this workshop at the Broad Institute originated in conversations about economic incentives, the exponential growth of multi-modal data sources, and challenging biomedical problems that have resulted in the life sciences emerging as both key consumers and producers of software and AI/ML technologies \cite{stephens2015big, ching2018opportunities, stateofai2020, natsecai}. We view this next decade as a critical growth phase for this process and an opportunity to shape the software engineering culture of the life sciences from the ground up. Safety and security, realized through both informal and formal methods, are central to this goal \cite{national2020safeguarding, fisher2017hacms, chong2016report}. \\

The result of these conversations was the event \href{http://fmie2021.github.io}{Formal Methods for the Informal Engineer} (FMIE), a workshop aimed at highlighting recent successes in the development of verified software. Given the relative youth of industrial-scale software engineering and distributed AI/ML platforms in the life sciences, we decided to make the content domain agnostic. Our aim was to expose researchers and developers in the life sciences to what was possible using formal methods in order to shape their outlook and guide future decision making. The workshop was a venue for collecting the success stories of formal methods, regardless of their domain, exploring applications to real-world software development, and connecting researchers to aspiring users.  \\

The event was organized as a ``flipped workshop,'' and on the first two days, to accommodate an audience consisting of many interested non-experts, we organized hands-on tutorials with two different verification technologies, the Coq and Z3 theorem provers \cite{coq_team, bertot2013interactive, de2008z3}. This provided many members of the audience with exposure to basic concepts of verification. The final day was organized as a traditional symposium with four sessions on General Topics, Verified Software Components, Distributed Systems, and Robust Machine Learning. \\

To more broadly communicate the lessons learned from organizing this workshop and the many conversations between a diverse group of life sciences researchers, physicians, software engineers, ML scientists, and programming language theorists, we make the following recommendations regarding the potential uses of and deployment strategies for formal methods. Although the workshop and these recommendations were conceived with a view towards organizations in the life sciences, they are domain agnostic and relevant to any organization that requires mission-critical or safety-critical software systems. \\

\pagebreak
\section{Recommendations}
\textbf{Recommendation 1:} \emph{Software leaders should investigate the use of formal methods as part of ongoing and future development.} \\
Formal methods are worth considering for many biomedical software projects, particularly those that require high assurances of data integrity, as in the case of patient data, that operate at scale and implement difficult concurrent algorithms, as in the case of aggregated genomic data, or that are safety-critical, such as medical devices in a hospital setting or automated, closed-loop biological laboratories. \\

\textbf{Recommendation 2:} \emph{Software leaders should ensure that their projects are held to the highest standards of quality and robustness from the standpoint of traditional software engineering prior to considering the use of formal methods. } \\
Formal methods are neither a substitute nor a cure for otherwise haphazard development practices.  Indeed, poor quality software foundations are one of the major hurdles to applying formal methods \cite{hinchey2008software}. Formal methods should be considered as an additional set of tools and development practices that can give assurances beyond what traditional methods can provide, but that rely on well-built software as a starting point \cite{newcombe2015amazon}. \\

\textbf{Recommendation 3:} \emph{Verification can be approached in an incremental manner, starting with the process of formalizing a specification.} \\
Investing in formal techniques does not need to be an all-or-nothing proposition. In particular, there is value in writing formal specifications for parts of a codebase, even if the results are never formally verified \cite{easterbrook1998formal, selsam2017developing}. The process of writing formal specifications alone can uncover bugs, especially architectural ones, early in the development process and result in many bugs being eliminated. Moreover, this process lays the foundation for a seamless transition to more rigorous verification when needed \cite{newcombe2015amazon, wayne2018practical, lamport1994introduction}.  \\

\textbf{Recommendation 4:} \emph{Software leaders should consider opportunities to build on top of already secure foundations.}\\
There are a number of existing software components, ranging from compilers, to microkernels, to file systems that have undergone rigorous verification and that could form the basis of new software in biomedicine \cite{fisher2017hacms, leroy2016compcert, klein2009sel4, chen2015using, tan2016new}. The purpose of Recommendation 5, embedding experts within existing teams, can aid the process of identifying appropriate foundations.  \\

\textbf{Recommendation 5:} \emph{Software leaders should consider embedding experts from the formal methods community within their software teams to assess how and where verification might be beneficial.} \\
Organizations should strive to build a rich culture of outstanding software engineering practices with a close link to formal methods. Even in situations where full-scale verification is not employed, FM tools such as static analyzers should be used routinely and architectural decisions should be informed by formal methods. Early education, project scoping, and strategic investment are key to ensuring that mission and safety-critical systems are available when needed.  Embedding formal methods experts within teams can play a critical role in identifying those tools, techniques, and training opportunities worth investing in and help to create a culture of safety that balances short and long-term needs. \\

\section{Conclusion}
The life sciences and medicine are poised to become leaders in the production of novel AI/ML technologies, many of which will be mission or safety-critical. Preparing for the widespread deployment of these systems is an urgent task, which demands long-term thinking and careful, deliberate attention to building bridges between communities with little previous interaction. Our experience with FMIE suggests that bringing together experts from the relevant fields of formal methods, programming language theory, and biomedicine can be done fruitfully and lead to concrete next actions to deepen the interaction between these spheres. We urge readers to consider the nexus of the life sciences and medicine with AI/ML and to work to ensure that the resulting technologies are developed not only with utmost regard to fairness and access, but also safety and security.  

\section*{Acknowledgements}
We would like to thank our sponsors Leaf Labs, Google DeepMind, CEA, and the Broad Institute of MIT and Harvard.  

\bibliographystyle{ieeetr}
\bibliography{fmie_workshop_recs}

\end{document}